\documentclass[12pt]{article}
\usepackage{amsmath}
\usepackage{amssymb}
\usepackage[dvips]{graphicx}
\usepackage[dvips]{epsfig}

 \advance \textwidth by 0.1cm
 \advance \textheight by 0.06cm
\def\##1{{\bf #1}}
\def\=#1{\underline{\underline #1}}
\def\4#1{\underline{\underline{\underline{\underline #1}}}}

\def\.{\mbox{ \tiny{$^\bullet$} }}

\def\le{\left(}
\def\ri{\right)}

\def\ric{\right\}}

\def\c#1{\cite{#1}}

\def\r#1{(\ref{#1})}

\def\epso{\epsilon_{\scriptscriptstyle 0}}
\def\lambdao{\lambda_{\scriptscriptstyle 0}}
\def\muo{\mu_{\scriptscriptstyle 0}}
\def\ko{k_{\scriptscriptstyle 0}}
\def\etao{\eta_{\scriptscriptstyle 0}}

\def\eps{\epsilon}
\def\epsa{\epsilon_a}
\def\epsb{\epsilon_b}
\def\epsc{\epsilon_c}

\def\epsmet{\eps_{met}}

\def\sp{\mathbf s}
\def\pinc{{\mathbf p}_+}
\def\pref{{\mathbf p}_-}

\def\Einc{{\mathbf E}_{inc}({\bf r})}
\def\Erefl{{\mathbf E}_{ref}({\bf r})}
\def\Etr{{\mathbf E}_{tr}({\bf r})}

\def\cpsi{\cos\psi}
\def\spsi{\sin\psi}
\def\ctheta{\cos\theta}
\def\stheta{\sin\theta}

\def\nr{n_\ell}

\def\rs{r_s}
\def\rp{r_p}
\def\ts{t_s}
\def\tp{t_p}

\def\ux{\hat{\mathbf{u}}_x}
\def\uy{\hat{\mathbf{u}}_y}
\def\uz{\hat{\mathbf{u}}_z}

\begin{document}

\noindent{\bf ENGINEERING THE PHASE SPEED OF
SURFACE--PLASMON WAVE AT THE PLANAR INTERFACE OF A METAL\\ AND
A CHIRAL SCULPTURED THIN FILM}

\vskip 18 pt

\noindent \textbf{Akhlesh Lakhtakia$^1$ and John A. Polo Jr$^2$}\\
\vskip0.1cm

\noindent $^1$CATMAS --- Computational \& Theoretical Materials Sciences Group,\\
\noindent Department of Engineering Science and Mechanics,\\
\noindent  Pennsylvania State University,\\
\noindent University Park, PA 16802, USA.\\
\noindent e--mail: akhlesh@psu.edu\\
\noindent fax: (814) 865 9974\\
\vskip 0.1cm
\noindent $^2$Department of Physics and Technology,\\
\noindent Edinboro University of Pennsylvania,\\
\noindent Edinboro, PA 16444, USA.\\
\noindent e--mail: polo@edinboro.edu\\
\noindent fax:  (814) 732 2455\\

\vskip 18pt

\noindent {\bf ABSTRACT} The solution of a boundary--value problem formulated for
a modified Kretschmann configuration shows that the phase speed of a surface--plasmon wave guided by the planar interface of a sufficiently thin metal film and a chiral sculptured thin film (STF) depends on the vapor incidence angle
used while fabricating the chiral STF by physical vapor depoistion. Therefore, it may be possible to engineer the
phase speed  quite simply by
selecting an appropriate value of the vapor deposition angle (in addition to the metal and the evaporant species).

\vskip 6pt


\section*{1. INTRODUCTION}

The solution of a boundary--value problem formulated for
a modified Kretschmann configuration has recently shown that a surface--plasmon wave can exist at the planar interface of a sufficiently thin metal film and a nondissipative structurally chiral medium,
provided the exciting plane wave is $p$--polarized \cite{LakhSPR}. Surface--plasmon waves
in the  visible and the near--infrared regimes
are being exploited nowadays for sensing, imaging, and other applications \cite{Ferrell,HYG99,BNC00,PSCE}. Structurally chiral mediums exist as
 chiral nematic and chiral smectic liquid
crystals  \cite{Chan}  as well as chiral sculptured thin films (STFs)
\cite{STFbook}.

In addition to their structural handedness (left or right) and their structural period along the axis of
helicoidal nonhomogeneity, chiral nematic liquid crystals are characterized by two relative permittivity
scalars: $\epsilon_a$ and $\epsilon_b$. Chiral smectic liquid crystals and chiral sculptured thin
films, however, are characterized by three relative permittivity scalars ($\epsilon_a$,
$\epsilon_b$, and $\epsilon_c$) as well as a tilt angle $\chi\in[0,\pi/2]$.  Our focus
here is on chiral STFs, which are
fabricated by physical vapor deposition \cite{STFbook}.

The constitutive parameters of a chiral STF can be substantially engineered by controlling a few
deposition variables.
While a chiral STF is being
grown on a planar substrate, its structural handedness
and structural period are controlled by substrate rotation, whereas a vapor incidence angle
$\chi_v\in(0,\pi/2]$ largely controls $\epsilon_{a,b,c}$ and $\chi$.
Therefore, it may be possible to engineer the phase speed of a surface--plasmon wave at the planar
interface of a thin metal film and a chiral STF in the Kretschmann configuration \cite{KR1968} quite
simply by selecting an appropriate value of $\chi_v$ (and the metal as well as the evaporant species, of
course).

That thought engendered this communication.
Section~2
contains a brief description of a modified Kretschmann configuration, wherein
the
combination of the metal
film and the chiral STF is sandwiched between two half--spaces occupied
by the same isotropic dielectric material that is optically denser than the chiral STF \cite{LakhSPR}.
Section~3 contains numerical results to
show that a higher value of the vapor incidence angle   reduces the phase speed of the surface--plasmon wave and increases the conversion efficiency of the
exciting plane wave into the surface--plasmon wave.

In the following sections, an $\exp(-i\omega t)$ time--dependence is implicit, with $\omega$
denoting the angular frequency. The free--space wavenumber, the
free--space wavelength, and the intrinsic impedance of free space are denoted by $\ko=\omega\sqrt{\epso\muo}$,
$\lambdao=2\pi/\ko$, and
$\etao=\sqrt{\muo/\epso}$, respectively, with $\muo$ and $\epso$ being  the permeability and permittivity of
free space. Vectors are in boldface, dyadics underlined twice;
column vectors are in boldface and enclosed within square brackets, while
matrixes are underlined twice and similarly bracketed. Cartesian unit vectors are
identified as $\ux$, $\uy$ and $\uz$.

\section*{2. THEORY IN BRIEF}\label{theory}
In conformance with the Kretschmann configuration for launching surface--plasmon waves,
the half--space $z\leq 0$ is occupied by a homogeneous, isotropic, dielectric
material described by the relative permittivity scalar $\eps_{\ell}$. Dissipation in this material
is considered to be negligible and its refractive index $n_\ell=\sqrt{\eps_\ell}$
is real--valued and positive. The laminar region
 $0 \leq z\leq L_{met}$ is occupied by a metal with relative permittivity
 scalar $\epsmet$.  Without significant
 loss of generality in the present context, the half--space
 $z\geq L_{met}+L_{stf}$ is taken to be occupied by the same
 material as fills the half--space $z\leq 0$.

 A chiral STF occupies the region
 $L_{met} \leq z \leq L_{met}+L_{stf}$. Its
relative permittivity dyadic $\=\epsilon_{stf}(z)$   is
factorable as
\begin{eqnarray}
\nonumber
&&
\=\epsilon_{stf}(z) =  \=S_z(z-L_{met})\.\=S_y(\chi)\.\=\epsilon^{
ref}_{stf}
\.\=S_y^T(\chi)\.\=S_z^T(z-L_{met})
\, , \\
&&\qquad\qquad\qquad L_{met}\leq z \leq L_{met}+L_{stf}\,,
\label{eps-chiSTF}
\end{eqnarray}
where  the reference relative permittivity dyadic
\begin{equation}
\=\epsilon_{stf}^{ref}= \epsa  \, \uz\uz  + \epsb \, \ux\ux
+ \epsc \, \uy\uy\,.
\end{equation}
The
dyadic function
\begin{equation}
\=S_z(z)=\le \ux\ux + \uy\uy \ri \cos{\le h\frac{\pi z}{\Omega} \ri}\\
+\le \uy\ux -
\ux\uy \ri \sin{\le h\frac{\pi z}{\Omega} \ri}+\uz\uz \, ,
\end{equation}
contains $2\Omega$ as the structural period and $h=\pm 1$ as the structural--handedness
parameter.  The tilt dyadic
\begin{equation}
\=S_y(\chi)=\le \ux\ux + \uz\uz \ri \cos{\chi}
+\le \uz\ux -
\ux\uz \ri \sin{\chi}+\uy\uy \,
\end{equation}
involves the angle   $\chi\in[0,\pi/2]$.
The superscript $^T$ denotes the transpose.

A $p$--polarized plane
wave, propagating in the half--space
$z \leq 0$ at an angle $\theta\in[0,\pi/2)$ to the $z$ axis and at an angle $\psi\in[0,2\pi)$
to the $x$ axis in the $xy$ plane, is incident on the metal--coated chiral STF.
The electric field phasor associated
with the incident plane wave is
\begin{equation}
\Einc=  \pinc \, e^{ i\kappa \le x\cpsi + y\spsi \ri
} \,e^{i\ko\nr z\ctheta}
\, , \qquad z \leq 0
\, .
\end{equation}
The reflected electric field phasor is expressed as
\begin{equation}
\Erefl= (\rs\,\sp +\rp \,\pref) \, e^{ i\kappa \le x\cpsi + y\spsi \ri
} \,e^{-i\ko\nr z\ctheta}
\, , \qquad z \leq 0
\, ,
\end{equation}
and the transmitted electric field phasor  as
\begin{equation}
\Etr= (\ts\,\sp +\tp\, \pinc) \, e^{ i\kappa \le x\cpsi + y\spsi \ri
} \,e^{i\ko\nr (z-L_{met}-L_{stf})\ctheta}
\, , \qquad z \geq L_{met}+L_{stf}
\, .
\end{equation}
Here,
\begin{equation}
\left.\begin{array}{l}
\kappa =
\ko\nr\stheta\\[5pt]
\sp=-\ux\spsi + \uy \cpsi
\\[5pt]
{\mathbf p}_\pm=\mp\le \ux \cpsi + \uy \spsi \ri \ctheta  + \uz \stheta
\end{array}\right\}
\, ,
\end{equation}
where $\omega/\kappa$ is the phase speed parallel to the interfacial
plane $z=L_{met}$ of interest, and the unit vectors $\#s$ and $\#p_{\pm}$ denote
the $s$-- and the $p$--polarization states of the electric field phasors.

The reflection amplitudes $\rs$ and $\rp$, as well as the transmission
amplitudes $\ts$ and $\tp$, have to be determined by the solution of
a boundary--value problem.
The required procedure  is standard
\cite[Chap. 10]{STFbook}, and is discussed in some detail
in the predecessor paper \cite{LakhSPR}. The quantity of interest is the absorbance
\begin{equation}
A_p=1-\left(\vert r_s\vert^2+\vert r_p\vert^2+\vert t_s\vert^2+\vert t_p\vert^2\right)\,.
\end{equation}

\section*{3. RESULTS AND DISCUSSION}\label{nrd}

Although chiral STFs may be made by evaporating a wide variety of materials \cite[Chap. 1]{STFbook}, the
constitutive parameters  of chiral STFs have not been extensively measured. However, the constitutive
parameters of certain columnar thin films (CTFs) are known.  Both CTFs and chiral STFs are fabricated by
physical vapor deposition. The basic procedure to deposit CTFs has been known for more than a century
\c{STFbook,SCW}. At low enough temperature and pressure, a solid material confined in a boat evaporates
towards a stationary substrate.  The vapor flux is collimated into a well--defined beam, and its average direction  is quantified by the angle
$\chi_v\in(0,\pi/2]$ with respect to the substrate plane.   Provided the adatom mobility is low, the
resulting film turns out to be an assembly of parallel and nominally identical nanorods \c{ZoneM}. The
nanorods have elliptical cross--sections and are tilted at an angle $\chi \geq \chi_v$ with respect to
the substrate plane. Equation \r{eps-chiSTF} applies for CTFs also, but after the limit
$\Omega\to\infty$ has been taken. The parameters $\eps_{a,b,c}$ and $\chi$ have to be functions of
$\chi_v$, at the very least because the nanoscale porosity of a CTF depends on the direction of the
vapor flux \c{KL}.

When the substrate is rotated about a normal passing through its centroid  at a constant angular velocity of reasonable
magnitude, parallel
nanohelixes grow instead of
parallel nanorods, and a chiral STF is deposited instead of a CTF \c{NO,MVS}. Although
the substrate is nonstationary, the functional relationships
connecting $\eps_{a,b,c}$ and $\chi$ to $\chi_v$ for CTFs would
substantially apply
for chiral STFs, since the vapor incidence angle $\chi_v$ remains
constant during the deposition of films of either kind.

A series of optical
characterization experiments on CTFs of the oxides of tantalum, titanium
and zirconium at $\lambdao=633$~nm were carried out
some years ago \c{HWH}. The results can be
put in the following form for the present purposes:
\begin{equation}
\label{Hodg1}
\left. \begin{array}{ll}
\eps_a  = \le n_{a0} + n_{a1}\,v + n_{a2}\,v^2\ri^2\\[5pt]
\eps_b = \le n_{b0} + n_{b1}\,v + n_{b2}\,v^2\ri^2\\[5pt]
\eps_c  = \le n_{c0} + n_{c1}\,v + n_{c2}\,v^2\ri^2\\[5pt]
\chi = \tan^{-1}\left(m\,\tan\chi_v\right)
\end{array}\ric\,.
\end{equation}
Here, $v =\chi_v/(\pi/2)$
is the vapor incidence angle expressed as a fraction of a right
angle.
 Table 1 contains values of the coefficients $n_{a0}$
to $m$    of CTFs of three different materials. We used these values to compute the numerical
results reported in this section. Let us note that although the bulk refractive indexes
of all three oxides are quite close to each other, the coefficients
$n_{a0}$
to $m$ of the three types of CTFs are quite different, as indeed are also their
constitutive parameters $\epsilon_{a,b,c}$ \cite{Chiadini}. These difference arise, in
part, due to the dependence of the
growth dynamics of a CTF on the evaporant species \cite{MVS}.

\vskip 10pt
\small{
\noindent{\bf Table 1.} Coefficients appearing
in Eqs. \r{Hodg1}, obtained \cite{Lpseudo}
from the experimental findings of Hodgkinson
{\em et al.\/} \cite{HWH}  on columnar thin films at $\lambdao=633$~nm. \\
\vskip 5pt

\begin{tabular}{|c||c|c|c||c|c|c||c|c|c||c|}
\hline
material & $n_{a0}$ & $n_{a1}$ & $n_{a2}$
              & $n_{b0}$ & $n_{b1}$ & $n_{b2}$
               & $n_{c0}$ & $n_{c1}$ & $n_{c2}$
              & $m$\\
\hline\hline
tantalum & 1.1961 & 1.5439 & $-$0.7719
               & 1.4600 & 1.0400 & $-$0.5200
               & 1.3532 & 1.2296 & $-$0.6148
        & 3.1056\\
oxide & & & & & &  & & &  & \\
\hline
titanium & 1.0443 & 2.7394& $-$1.3697
               & 1.6765 & 1.5649 & $-$0.7825
               & 1.3586 & 2.1109 & $-$1.0554
        & 2.8818\\
oxide & & & & & &  & & &  & \\
\hline
zirconium & 1.2394 & 1.2912& $-$0.6456
               & 1.4676 & 0.9428 & $-$0.4714
               & 1.3861 & 0.9979 & $-$0.4990
        & 3.5587\\
oxide & & & & & &  & & &  &  \\
\hline

\end{tabular}
}
\vskip 10pt


\begin{figure}[!htb]
\centering \psfull
\epsfig{file=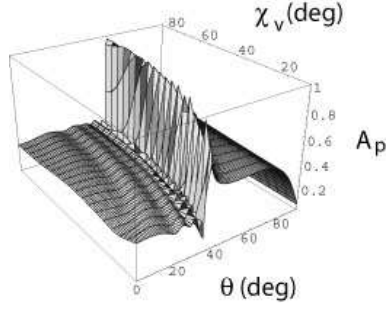,
width=5cm}
\caption{Absorbance $A_p$  as a function of $\chi_v$ and $\theta$ when $\psi=0^\circ$,
$\lambdao=633$~nm,
and the incident plane wave is $p$--polarized. The chiral STF is made of tantalum
oxide; see Eqs.~\r{Hodg1}
and Table 1 for $\eps_{a,b,c}$ and $\chi$ as functions of $\chi_v$.
Other parameters are
as follows:  $\Omega=200$~nm, $h=\pm1$,
$L_{stf}=4\Omega$,
$\epsmet=-56+i21$, $L_{met}=15$~nm, and
$\eps_\ell=6.656$.
}
\label{figTaO}
\end{figure}


As can be gleaned from the predecessor paper \cite{LakhSPR}, just a 2--period thick chiral STF should
suffice for the excitation of a surface--plasmon wave; hence, we set $L_{stf}=4\Omega$ with
$\Omega=200$~nm. The absorbance $A_p$ was calculated as a function of both $\theta$ and $\chi_v$ at the
free--space wavelength $\lambdao=633$~nm, for both structurally left--handed
($h=-1$) and structurally
right--handed ($h=+1$) chiral STFs. The metal was chosen to be aluminum: $\epsmet=-56+i21$ and
$L_{met}=15$~nm. The two half--spaces were taken to be filled with zinc selenide
($\eps_\ell=6.656$), which is optically denser than all three types of chiral STFs
considered here.

\begin{figure}[!htb]
\centering \psfull
\epsfig{file=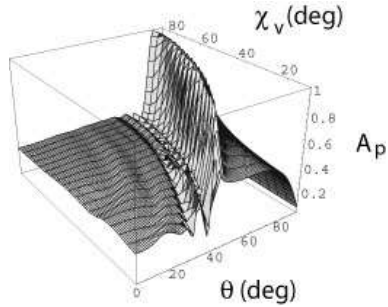,
width=5cm}
\caption{Same as Fig.~\ref{figTaO}, except that the chiral STF is made of
titanium oxide.
}
\label{figTiO}
\end{figure}

\begin{figure}[!htb]
\centering \psfull
\epsfig{file=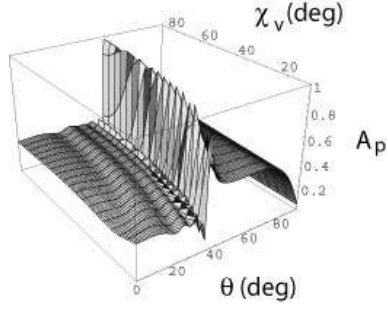,
width=5cm}
\caption{Same as Fig.~\ref{figTaO}, except that the chiral STF is made of
zirconium oxide.
}
\label{figZrO}
\end{figure}

Figures \ref{figTaO}, \ref{figTiO}, and \ref{figZrO} show $A_p$ as a function of $\theta$ for different
values of $\chi_v$ for  chiral STFs made of tantalum oxide, titanium oxide, and zirconium oxide,
respectively. For low values of $\theta$, $A_p$ is small ($\lesssim 0.4$) in all three figures. As
$\theta$ increases for a fixed value of $\chi_v$, $A_p$ begins to fluctuate, then records a very high
value, and then tends to zero as $\theta\to \pi/2$. The sharp $A_p$--peak indicates the excitation of a
surface--plasmon wave at the interface of the metal and the chiral STF
\cite{LakhSPR}.


\begin{figure}[!htb]
\centering \psfull
\epsfig{file=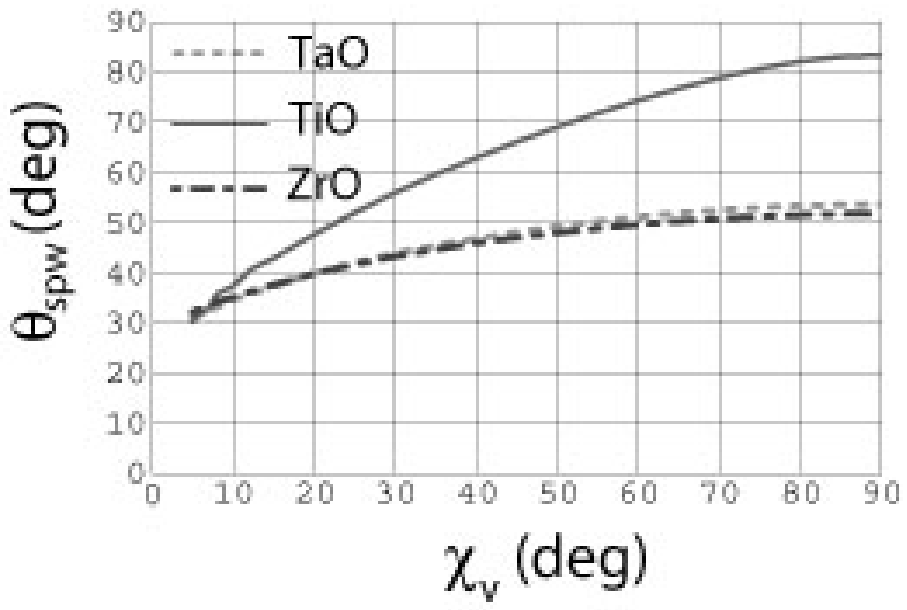,
width=5cm}
\caption{$\theta_{spw}$ as a function of $\chi_v$, when the chiral  STF is made of tantalum oxide, titanium oxide, or zirconium oxide. Other parameters are
as follows:  $\lambdao=633$~nm, $\psi=0$, $\Omega=200$~nm, $h=\pm1$,
$L_{stf}=4\Omega$,
$\epsmet=-56+i21$, $L_{met}=15$~nm, and
$\eps_\ell=6.656$.
}
\label{figTheta-spw}
\end{figure}



\begin{figure}[!htb]
\centering \psfull
\epsfig{file=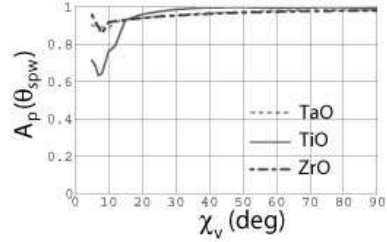,
width=5cm}
\caption{$A_p$ at $\theta=\theta_{spw}$ as a function of $\chi_v$, when the chiral  STF is made of tantalum oxide, titanium oxide, or zirconium oxide. Other parameters are
as follows:  $\lambdao=633$~nm, $\psi=0$, $\Omega=200$~nm, $h=\pm1$,
$L_{stf}=4\Omega$,
$\epsmet=-56+i21$, $L_{met}=15$~nm, and
$\eps_\ell=6.656$.
}
\label{figAp-spw}
\end{figure}


Let $\theta_{spw}$ denote the value of $\theta$ at which the surface--plasmon wave
is excited. Figure~\ref{figTheta-spw} contains plots of $\theta_{spw}$ versus $\chi_v$
gleaned from the previous three figures. Clearly, as $\chi_v$ increases, so does
$\theta_{spw}$ for chiral STFs fabricated with a specific evaporant species. But the wavenumber of the surface--plasmon wave is given by
\begin{equation}
\kappa_{spw}=\ko n_\ell\,\sin\theta_{spw}
\,.
\end{equation}
Hence, $\kappa_{spw}$ is a monotonically increasing function of $\chi_v$, which means that the phase speed
\begin{equation}
v_{spw}=\omega/\kappa_{spw}
\end{equation}
is a monotonically decreasing function of $\chi_v$.  This conclusion agrees qualitatively with
earlier studies wherein CTFs were considered in lieu of chiral STFs \cite{LPajp,PLmotl}.

Figure~\ref{figAp-spw} contains plots of $A_p$ at $\theta=\theta_{spw}$ versus $\chi_v$ gleaned from
Figs.~\ref{figTaO}--\ref{figZrO}. As $\chi_v$ increases, the conversion of the incident energy into the
energy of the surface--plasmon wave becomes more efficient after an initial drop at very low $\chi_v$. There is, however, a threshold value of
$\chi_v$ beyond which the conversion efficiency does not increase appreciably.

Although the numerical results presented in Figs.~\ref{figTaO}--\ref{figAp-spw} hold strictly
for $\psi = 0$, we have verified that they do not change substantially for other values of
$\psi$. Furthermore, the choice of evaporant species affects the characteristics of the surface--plasmon wave, as is evident from the significantly different graphs in Figs.~\ref{figTheta-spw} and \ref{figAp-spw} for
titanium--oxide films than for the other two types of films. This is not surprising because the relative permittivity scalars $\epsilon_{a,b,c}$ of the zirconium--oxide and
tantalum--oxide films are close to each other but quite different from those of the
titanium--oxide films \cite{Chiadini}.

In conclusion, selection of a higher vapor incidence angle leads to a lower phase speed of the surface--plasmon wave at the
planar interface of a metal film and the chiral sculptured thin film, in the Kretschmann
configuration.  Concurrently, the conversion efficiency of the exciting plane wave into the
surface--plasmon wave is higher. Thus, the characteristics of
the surface--plasmon wave can be substantially engineered by the proper selection of the vapor incidence
angle.

\newpage

\end{document}